\documentclass[12pt]{article}
\usepackage{epsfig,times,helvet,exscale,latexsym,psfrag}
\DeclareFontFamily{U}{msb}{}
\DeclareFontShape{U}{msb}{m}{n}{
<5><6><7><8><9> gen *msbm <10><10.95><12><14.4><17.28><20.74><24.88>msbm10}{}
\DeclareSymbolFont{AMSb}{U}{msb}{m}{n}
\DeclareMathSymbol{\bA}{\mathbin}{AMSb}{'101}
\DeclareMathSymbol{\bB}{\mathbin}{AMSb}{'102}
\DeclareMathSymbol{\bC}{\mathbin}{AMSb}{'103}
\DeclareMathSymbol{\bD}{\mathbin}{AMSb}{'104}
\DeclareMathSymbol{\bE}{\mathbin}{AMSb}{'105}
\DeclareMathSymbol{\bF}{\mathbin}{AMSb}{'106}
\DeclareMathSymbol{\bG}{\mathbin}{AMSb}{'107}
\DeclareMathSymbol{\bH}{\mathbin}{AMSb}{'110}
\DeclareMathSymbol{\bI}{\mathbin}{AMSb}{'111}
\DeclareMathSymbol{\bJ}{\mathbin}{AMSb}{'112}
\DeclareMathSymbol{\bK}{\mathbin}{AMSb}{'113}
\DeclareMathSymbol{\bL}{\mathbin}{AMSb}{'114}
\DeclareMathSymbol{\bM}{\mathbin}{AMSb}{'115}
\DeclareMathSymbol{\bN}{\mathbin}{AMSb}{'116}
\DeclareMathSymbol{\bO}{\mathbin}{AMSb}{'117}
\DeclareMathSymbol{\bP}{\mathbin}{AMSb}{'120}
\DeclareMathSymbol{\bQ}{\mathbin}{AMSb}{'121}
\DeclareMathSymbol{\bR}{\mathbin}{AMSb}{'122}
\DeclareMathSymbol{\bS}{\mathbin}{AMSb}{'123}
\DeclareMathSymbol{\bT}{\mathbin}{AMSb}{'124}
\DeclareMathSymbol{\bU}{\mathbin}{AMSb}{'125}
\DeclareMathSymbol{\bV}{\mathbin}{AMSb}{'126}
\DeclareMathSymbol{\bW}{\mathbin}{AMSb}{'127}
\DeclareMathSymbol{\bX}{\mathbin}{AMSb}{'130}
\DeclareMathSymbol{\bY}{\mathbin}{AMSb}{'121}
\DeclareMathSymbol{\bZ}{\mathbin}{AMSb}{'132}

\def\DK{\varepsilon}
\def\nn{{\bf n}}
\def\rr{{\bf r}}

\def\DD{{\bf D}}
\def\prt{\partial}
\def\ch#1{\chi\rule[-1ex]{0mm}{2ex}_{#1}}
\def\avg#1{\overline{#1}}
\def\DKe{\avg{\DK}}
\def\phe{\avg{\phi}}

\begin{document}
\vspace*{-2cm}{\tt to appear in: Granular Matter (1999), in print}
\begin{center}
{\large\bf MACROSCOPIC DIELECTRIC CONSTANT FOR MICROSTRUCTURES
OF SEDIMENTARY ROCKS 
\footnote{Paper also presented at Third Workshop on
``Electromagnetic Wave Interaction with Water and Moist Substances'',
Athens, Georgia, April 12, 1999}
}\\[24pt]
{\em R. Hilfer$^{1,2}$, J. Widjajakusuma$^1$ and B. Biswal$^{1,3}$}\\[24pt]
$~^1$ICA-1, Universit{\"a}t Stuttgart, 70569 Stuttgart, Germany\\
$~^2$Institut f{\"u}r Physik, Universit{\"a}t Mainz, 55099 Mainz, Germany\\
$~^3$Department of Physics \& Electronics, Sri Venkateswara College, \\
University of Delhi, New Delhi - 110 021, India\\\end{center}
{\em\bf ABSTRACT.}
An approximate method to calculate dielectric response
and relaxation functions for water saturated sedimentary
rocks is tested for realistic threedimensional pore
space images.
The test is performed by comparing the prediction
from the approximate method against the exact solution.
The approximate method is based on image analysis and 
local porosity theory.
An empirical rule for the specification of the
length scale in local porosity theory is advanced.
The results from the exact solution are compared
to those obtained using local porosity theory and
various other approximate mixing laws.
The calculation based on local porosity theory is found 
to yield improved quantitative agreement with the exact 
result.\\[24pt]

\section*{\normalsize\rm 1. INTRODUCTION}
Applying a small electric field to a heterogeneous mixture of 
homogeneous and isotropic dielectrics can give rise to an effective 
dielectric behaviour that may differ substantially from
that of the constituents.
The effective dielectric constant of the mixture depends 
not only on the dielectric constants of the constituent 
materials, but also on the geometrical microstructure of 
the mixture \cite{bot73,BB78}.
It is tempting to utilize the dependence on the microstructure for 
deducing microstructural information from dielectric measurements.
This motivates (at least partially)  the continued
interest in the dielectric response of waterfilled sedimentary 
rocks or soils \cite{kor84,ken84,KN87,HN89,TKS90,has95,fri98}.
Such studies are important for the interpretation of geophysical 
or petrophysical borehole measurements \cite{PNW78,HN85}.
Before microstructural information can be deduced 
from dielectric measurements a reliable theory is needed 
that links the observed dielectric response with the desired 
microstructural information.
Our objective in this paper is to discuss 
local porosity theory \cite{hil91d} as an approximate
relation between microstructural information and dielectric response.
We compare its prediction against classical mixing laws and against
the exact value for the effective dielectric constant.
Our discussion starts with reminding the reader how
the microstructure enters via the microscopic equations.
Next we review briefly the most popular mixing laws.
In Section 4 we recall the basics of local porosity theory ,
and advance a new length scale, called the percolation length.
In Section 5 we present results for four different threedimensional
sandstone samples.
We give only a short summary of selected results.
A more detailed account is in preparation \cite{WBH99}.

\section*{\normalsize\rm 2. MICROSCOPIC EQUATIONS}
Microscopically the electric fields and potentials are governed 
by Maxwells equations in the quasistatic approximation.
To be more specific let us consider from now on a two phase
mixture of water and rock.
The water fills the pore space of the rock.
The pore space will be denoted as $\bP$, while the rock matrix
will be denoted as $\bM$.
The sets $\bP,\bM\in\bR^3$ are subsets of threedimensional
space, and their union $\bS=\bP\cup\bM$ represents the sample.
Within the quasistatic approximation the electrical potential $U(\rr)$ 
obeys the equation
\begin{eqnarray}
\nabla\cdot\DD(\rr) & = & 0 , \hspace*{4cm}\rr\in\bS,\rr\notin\prt\bM\\
\label{maxwell}
\DD(\rr) & = & -\DK(\rr)\nabla U(\rr) , \hspace*{2cm}\rr\in\bS,\rr\notin\prt\bM
\label{constlaw}
\end{eqnarray}
where $\DD(\rr)$ is the electric displacement vector,
\begin{equation}
\DK(\rr) = \DK_\bP\ch{\bP}(\rr)+\DK_\bM\ch{\bM}(\rr)
\label{DF}
\end{equation}
is the inhomogeneous dielectric function, and $\prt\bM=\prt\bP$ 
denotes the internal boundary between the two phases.
Here
\begin{equation}
\ch{\bG}(\rr) = \left\{
\begin{array}{r@{\quad:\quad}l}
1 & \mbox{for}\quad \rr\in\bG \\
0 & \mbox{for}\quad \rr\notin\bG
\end{array}
\right .
\label{charfunc}
\end{equation}
denotes the indicator function of a set $\bG$,
and $\DK_\bP,\DK_\bM$ are the dielectric constants
(possibly frequency dependent) of the constituents 
(water and rock).
For real rocks with typical pore sizes of order 
$100\mu$m the quasistatic approximation
remains valid for frequencies up to several $100$GHz.

Differentiation of the dielectric function $\DK(\rr)$
is not allowed at the internal boundary $\prt\bM=\prt\bP$,
and hence the equations must be supplemented with 
boundary conditions on the internal boundary.
Assuming that there is no surface charge density,
and denoting the unit normal to the
surface by $\nn$ we have
\begin{eqnarray}
\lim_{\eta\to 0} \nn\cdot \DD(\rr+\eta \nn)
&=& \lim_{\eta\to 0} \nn\cdot \DD(\rr-\eta \nn), \hspace*{1.5cm}\rr\in\prt\bM
\label{bc1}\\
\lim_{\eta\to 0} \nn\times \nabla U(\rr+\eta \nn) &=&
\lim_{\eta\to 0} \nn\times \nabla U(\rr-\eta \nn), \hspace*{1cm}\rr\in\prt\bM
\label{bc2}
\end{eqnarray}
where the second condition expresses continuity of the
electric field component tangential to the surface.
These equations must be supplemented with further
boundary conditions representing the applied
external potential at the sample boundaries $\prt\bS$.

Given a solution $U(\rr;\prt\bM)$ of the above equations
all macroscopic properties of interest can in principle
be calculated from it.
An example would be the effective macroscopic dielectric
constant $\DKe$.
It can be defined by averaging the solution
$U(\rr;\prt\bM)$ of the microscopic problem
\begin{equation}
\avg{\DD(\rr)} =  -\DKe\;\avg{\nabla U(\rr)}
\label{eff}
\end{equation}
where $\avg{A}$
formally represents a suitably defined ensemble or 
spatial average of a quantity $A$ over microstructures.
The corresponding measure on the space of microstructures
generally depends on parameters $\pi_i$ characterizing the
statistical properties of the microstructure.
In practice the microstructure
is usually not known in sufficient detail.
As a consequence it is impossible to calculate
$U(\rr;\prt\bM)$, and one has to resort to 
approximate theories based on incomplete knowledge.
This gives rise to the so called mixing laws.
Mixing laws provide a relationship between the dielectric 
constants $\DK_i$ of the constituent phases $i=1,2,...$ 
and the effective macroscopic dielectric constant of the
mixture $\DKe$. 
They have the general form
\begin{equation}
\DKe = \DKe(\DK_1,\DK_2,...;\pi_1,\pi_2,...)
\end{equation}
where $\pi_i$ are parameters reflecting the 
dependence on the microstructure.
These parameters determine which information about the
microstructure can be obtained from measurements
of the effective dielectric constant.

\section*{\normalsize\rm 3. MIXING LAWS AND RIGOROUS BOUNDS}
Let us illustrate the approximations involved in deriving
mixing laws with a simple example. 
If we define a spatial average over the whole sample as
\begin{equation}
\avg{f} = \frac{1}{|\bS|}\int f(\rr)\ch{\bS}(\rr)d^3\rr
\label{average}
\end{equation}
where
\begin{equation}
|\bG|=\int\ch{\bG}(\rr)d^3\rr
\end{equation}
denotes the volume of a set $\bG\in\bR^3$,
then we obtain by inserting eq. (\ref{DF}) into 
eq. (\ref{constlaw}) and applying
eq. (\ref{average}) 
\begin{equation}
\avg{\DD} =  -\DK_\bP\;\avg{\ch{\bP}\nabla U}
-\DK_\bM\;\avg{\ch{\bM}\nabla U}
\end{equation}
If we assume $\avg{\ch{\bP}\nabla U}=\avg{\ch{\bP}}\;\;\avg{\nabla U}$
in the spirit of mean field theories, then we obtain by comparison 
with eq. (\ref{eff}) the mixing law of the arithmetic average
\begin{equation}
\DKe = \phe\DK_\bP+(1-\phe)\DK_\bM
\label{AA}
\end{equation}
where
\begin{equation}
\phe = \frac{|\bP|}{|\bS|}
\label{porosity}
\end{equation}
denotes the porosity, i.e. the volume fraction of pore space.

Other popular mixing laws for homogeneous and
isotropic systems, similar to eq. (\ref{AA}),
include the harmonic averages
\begin{equation}
\DKe = \left(\phe\DK^{-1}_\bP+(1-\phe)\DK^{-1}_\bM\right)^{-1} ,
\label{HA}
\end{equation}
the Clausius-Mossotti approximation with $\bP$ as background phase
\begin{equation}
\DK_C(\phe) = \DK_\bP\left( 1 -
\frac{1-\phe}{(1-\DK_\bM/\DK_\bP)^{-1}-\phe /3} \right) =
\DK_\bP\left(\frac{3\DK_\bM+2\phe(\DK_\bP-\DK_\bM)}
{3\DK_\bP-\phe(\DK_\bP-\DK_\bM)}\right),
\label{epsc} 
\end{equation}
the Clausius-Mossotti approximation with $\bM$ as background phase
\begin{equation}
\DK_B(\phe) = \DK_\bM\left( 1 -
\frac{\phe}{(1-\DK_\bP/\DK_\bM)^{-1}-(1-\phe)/3} \right) =
\DK_\bM\left(\frac{2\DK_\bM+\DK_\bP+2\phe(\DK_\bP-\DK_\bM)}
{2\DK_\bM+\DK_\bP-\phe(\DK_\bP-\DK_\bM)}\right),
\label{epsb}
\end{equation}
and the self-consistent effective medium approximation
\begin{equation}
\phe\frac{\DK_\bP-\avg{\DK}}{\DK_\bP+2\avg{\DK}} + 
(1-\phe)\frac{\DK_\bM-\avg{\DK}}{\DK_\bM+2\avg{\DK}} = 0 
\label{EMA}
\end{equation}
which leads to a quadratic equation for $\DKe$.
In all of these mixing laws the porosity $\phe$ is all that
is left to characterize the microstructure.
Measurement of $\DKe$ combined with the knowledge of $\DK_\bM,\DK_\bP$
allows to deduce the porosity from such formulae.
It is clear, however, that only the value for $\phe$ given in
eq. (\ref{porosity}) is correct, and this raises the issue
of how accurate the approximate mixing laws actually are.

In such a situation it is useful to have rigorous upper and lower 
bounds for $\DKe$.
Such bounds can be derived by variational methods under
various assumptions about the stochastic nature of the microstructure
\cite{HS62,ber82,tor91}.
If the microstructure is known to be homogeneous and isotropic with
bulk porosity $\phe$, and if $\DK_\bP > \DK_\bM$, then 
\begin{equation}
\DK_B(\phe) \leq \DKe \leq \DK_C(\phe)
\end{equation}
holds, where the upper and the lower bound are given by
the Clausius-Mossotti formulae, eqs. (\ref{epsb}) and
(\ref{epsc}).
For $\DK_\bP < \DK_\bM$ the bounds are reversed.
It turns out that the simplest mixing laws, eq. (\ref{AA}) 
and (\ref{HA}), violate these bounds.
They will not be considered further.

\section*{\normalsize\rm 4. LOCAL POROSITY THEORY}
A fundamental drawback of the classical bounds and mixing laws
is that they depend only on porosity $\phe$ as a single geometric
parameter characterising the complex microstructures.
Recently a new mixing law was developed which circumvents this
restriction by incorporating fluctuations in porosity and connectivity
\cite{hil91d,hil92a,hil94b,hil94g,hil95d,hil98a}.
The basic idea of the new approach, called local porosity theory,
is to measure fundamental geometric observables (such as
volume fraction, surface density, mean curvature density,
Euler-characteristic or connectivity) within a bounded (compact) 
subset of the porous medium and to collect these measurements
into various histograms.
These histograms are then used in a generalization of
the effective medium approximation to predict effective
transport properties.

Let $\bK(\rr,L)\in \bS$ denote a cube of sidelength $L$
centered at $\rr$.
The set $\bK(\rr,L)$ defines a measurement cell (window)
inside of which local geometric properties such
as porosity or specific internal surface are
measured \cite{hil91d}.
The local porosity in this measurement cell $\bK(\rr,L)$ 
is defined as
\begin{equation}
\phi(\rr,L)=\frac{|\bP\cap\bK(\rr,L)|}{|\bK(\rr,L)|}
\label{lpd1}
\end{equation}
The local porosity distribution $\mu(\phi,L)$ is defined as
\begin{equation}
\mu(\phi,L) = \frac{1}{m}\sum_\rr\delta(\phi-\phi(\rr,L))
\label{lpd2}
\end{equation}
where $\delta(x)$ denotes Diracs $\delta$-distribution, and
the summation runs over placements of the measurement cell.
The integer
\begin{equation}
m = \prod^3_{i=1}(M_i-L+1) 
\label{lpd3}
\end{equation}
is the number of placements of $\bK(\rr,L)$ for a discretized
sample (assumed to be a parallelepiped) with sidelengths $M_i$.
Ideally all measurement cells should be disjoint \cite{hil91d}, 
but in practice this would give very poor statistics.
The support of $\mu(\phi,L)$ is the unit interval $0\leq\phi\leq 1$
for all $L$.

The second geometrical ingredient for local porosity theory 
characterizes the connectivity of each measurement cell.
We define
\begin{equation}
\Lambda_x(\rr,L)=\left\{
\begin{array}{r@{\quad:\quad}l}
1 & {\rm if~}\prt\bK_{-x}(\rr,L)\leadsto \prt\bK_{+x}(\rr,L)
{\rm ~in~}\bK(\rr,L)\cap\bP \\[6pt]
0 & {\rm otherwise}
\end{array}
\right.
\label{lpp1}
\end{equation}
where $\leadsto$ indicates that there exists a path inside the pore space
of the measurement cell
which connects the left boundary $\prt\bK_{-x}(\rr,L)$ of $\bK(\rr,L)$
perpendicular to the $x$-direction to its right boundary 
$\prt\bK_{+x}(\rr,L)$.
Similarly we define $\Lambda_y(\rr,L)$ for percolation in the
$y$-direction, and $\Lambda_z(\rr,L)$ for the $z$-direction.
It is possible to relate these quantities to the Euler-characteristic
of $\bP$  \cite{hil99}.
Given eq. (\ref{lpp1}) we define the local percolation probability
\begin{equation}
\lambda(\phi,L) = \frac{
\sum_\rr \Lambda_x(\rr,L)\Lambda_y(\rr,L)\Lambda_z(\rr,L)
\delta_{\phi\phi(\rr,L)}}
{\sum_\rr\delta_{\phi\phi(\rr,L)}} 
\label{lpp2}
\end{equation}
which gives the probability that a cell with local porosity
$\phi$ percolates in all three directions.

With these preparations the mixing law of local porosity 
theory reads \cite{hil91d}
\begin{equation}
\int_0^1\!\!\frac{\DK_C(\phi)-\DKe}
{\DK_C(\phi)+2\DKe}\lambda(\phi,L)\mu(\phi,L)d\phi +
\int_0^1\!\!\frac{\DK_B(\phi)-\DKe}
{\DK_B(\phi)+2\DKe}(1-\lambda(\phi,L))\mu(\phi,L)d\phi = 0
\label{LPT1}
\end{equation}
where $\DK_B$ and $\DK_C$ are given in eqs. (\ref{epsb}) and
(\ref{epsc}), $\mu$ in eq. (\ref{lpd2}), and $\lambda$ in
eq. (\ref{lpp2}).

The mixing law (\ref{LPT1}) is a generalization of the 
effective medium approximation.
In fact, it reduces to eq. (\ref{EMA}) in the limit $L\to 0$.
In the limit $L\to\infty$ it also reduces to eq. (\ref{EMA})
albeit with $\phe$ in eq. (\ref{EMA}) replaced with $\lambda(\phe)$.
In both limits the basic assumptions underlying all effective
medium approaches become invalid.
For small $L$ the local geometries become strongly 
correlated, and this is at variance with the basic
assumption of weak or no correlations.
For large $L$ on the other hand the assumption that
the local geometry is sufficiently simple becomes
invalid \cite{hil91d}.
Hence we expect that formula (\ref{LPT1}) will yield 
good results only for intermediate $L$.

The question which $L$ to choose has been discussed
in the literature \cite{hil92f,hil96g,hil98a,hil98h}.
Here we advance a new proposal.
We suggest to use a length scale $L_p$, called
the percolation length.
It is defined using the function 
\begin{equation}
p(L) = \int_0^1 \mu(\phi,L)\lambda(\phi,L)\;d\phi
\label{lpp3}
\end{equation}
which gives the total fraction of percolating cells
at length $L$.
Experience shows that $p(L)$ frequently has a sigmoidal
shape, and this has led us to define $L_p$ as the length
scale corresponding to the inflection point of $p(L)$.
Hence we define $L_p$ through the condition
\begin{equation}
\left.\frac{d^2 p}{dL^2}\right|_{L=L_p}=0
\label{lpp4}
\end{equation}
assuming that it is unique.
The idea behind this definition is that at the inflection
point the function $p(L)$ changes most rapidly from its
trivial value $p(0)=\phe$ at small $L$ to its equally 
trivial value $p(\infty)=1$ at large $L$ (assuming that the
pore space percolates).
We have observed that the length $L_p$ is typically much
larger than the correlation length \cite{hil98a,BMHSO99}.

We remark that there are two other important length scales
associated with $p(L)$.
The first of these is the threshold length $L_c$ defined
by 
\begin{equation}
p(L_c)=p_c
\end{equation}
where $p_c$ can be taken as
the percolation threshold for the underlying lattice
($p_c\approx 0.248812$ for the simple cubic lattice
\cite{hug96})
or as $p_c=1/3$ for the effective medium approximation.
This length scale is particularly important for
network models which attempt to replace the 
complex microstructure by an effective lattice
with similar statistical properties as the real sample.
Of course $L_c$ may not exist when $\phe=p(0)>p_c$.

The second length scale $L_\delta$ is the length at which 
$p(L)$ approaches its asymptotic value $p(\infty)$
to a given degree of accuracy.
For percolating samples $p(\infty)=1$.
We define $L_\delta$ through
\begin{equation}
|p(\infty)-p(L_\delta)|<\delta
\end{equation}
for small $\delta>0$.
The length $L_\delta$ may be equated with the size of
the so called ``representative elementary volume'' (REV)\cite{bea72}
required for representativity with respect to connectivity.
$L_\delta$ represents the scale of the averaging (smoothing) region
that is needed to ensure that the fluctuating microscopic connectivity
can be replaced with an averaged connectivity field defined on the continuum.
The small parameter $\delta$ controls the degree of smoothness.
Naturally we expect $L_p<L_\delta$ for small enough $\delta$.

\section*{\normalsize\rm 5. RESULTS}
We have analyzed four samples of sedimentary rocks whose
pore spaces were obtained by computer assisted microtomography.
Images of two of the samples are given in Figures 1 and 3 of
\cite{hil98a}.
Each data set consists of a threedimensional array of 0's 
and 1's indicating pore space $\bP$ or matrix $\bM$.
The dimensions of the array are $M_1,M_2$ and $M_3$.
Table I gives a synopsis of the characteristics 
of the four samples.
\begin{center}
{TABLE I: Overview over properties of the data sets
for four reservoir sandstones.}\\[8pt]
\begin{tabular*}{\textwidth}{|l@{\extracolsep{\fill}}llllll|} \hline
Sample & Description & $a$ & $M_1\times M_2\times M_3$ & $L_p$ &
$\phe$ & $\DKe$ \\\hline
A & Berea & $10\mu$m & $128\times 128\times 128$ & $180\mu$m &
$0.1775$ & $9.827$ \\
B & coarse Sst20d & $30\mu$m & $73\times 128\times 128$ & $420\mu$m &
$0.2470$ & $13.073$ \\
C & fine Sst6d & $10\mu$m & $95\times 128\times 128$ & $100\mu$m &
$0.3200$ & $16.934$ \\
D & Fontainebleau & $7.5\mu$m & $300\times 300\times 299$ & $225\mu$m &
$0.1355$ & $8.599$ \\
\hline
\end{tabular*}
\end{center}
Here $a$ is the resolution, and $M_i$ are the dimensionless sidelengths
of the sample in units of $a$.
The bulk porosity $\phe$ was defined in eq. (\ref{porosity}) and the length
$L_p$ in eq. (\ref{lpp4}).

We have solved the microscopic equations (\ref{maxwell})-(\ref{bc2})
numerically using the values $\DK_\bP\approx 87.74$ 
and $\DK_\bM=4.7$ (in units of $\DK_0=8.854\cdot 10^{-12}$F/m) for the 
dielectric constants, and calculated $\DKe$ from
equation (\ref{eff}) using the averaging procedure defined 
in eq. (\ref{average}).
The results are shown in the last column in Table I.

\begin{center}
\epsfig{file=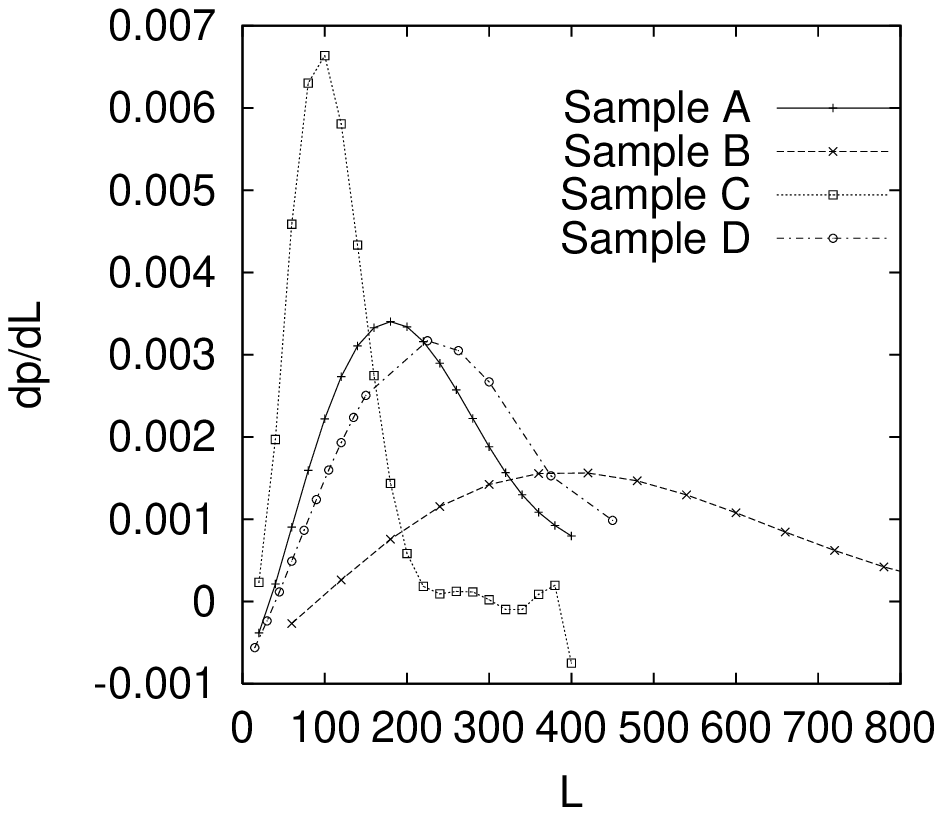,width=135mm}\\
Figure 1: 
Derivative of $p(L)$ (in $\mu$m$^{-1}$) defined in eq. (\ref{lpp3}).
The abscissa is the size of measurement cells in $\mu$m.
The percolation length $L_p$ corresponds to the position 
of the maximum.
\end{center}

Next we measured $\mu(\phi,L)$ and $\lambda(\phi,L)$
as a function of $\phi$ and $L$.
For the determination of $\lambda$ we used a Hoshen-Kopelman
algorithm \cite{SA92}.
Integrating the product according to eq. (\ref{lpp3})
and differentiating the result with respect to $L$ we
find the curves shown in Figure 1.
The locations of the maxima give the values of $L_p$ tabulated
in Table I.

Finally we solve equation (\ref{LPT1}) iteratively to find the
value of $\DKe_{LPT}$ predicted by local porosity theory.
We have plotted these values together with the predictions
from the other mixing formulae in Figure 2.
We emphasize that, contrary to spectral theories
or network models,  neither the mixing laws nor the local
porosity theory contains any free fitting parameters.
While the Clausius-Mossotti predictions (upper and lower
bounds) do not give good estimates
the results from the effective medium approximation
and the local porosity theory are in much better 
agreement with the exact result.
Note however that the effective medium values 
approach zero for infinite contrast, i.e. $\DKe_{EMA}\to 0$ for
$\DK_\bP/\DK_\bM\to\infty$.
The values of $\DKe_{LPT}$ on the other hand remain finite
and are in similarly good agreement also in that limit \cite{WBH99,hil98h}.

\begin{center}
\psfrag{X}{\Huge $\bar{\phi}$}
\psfrag{Y}{\Huge $\bar{\epsilon}$}
\epsfig{file=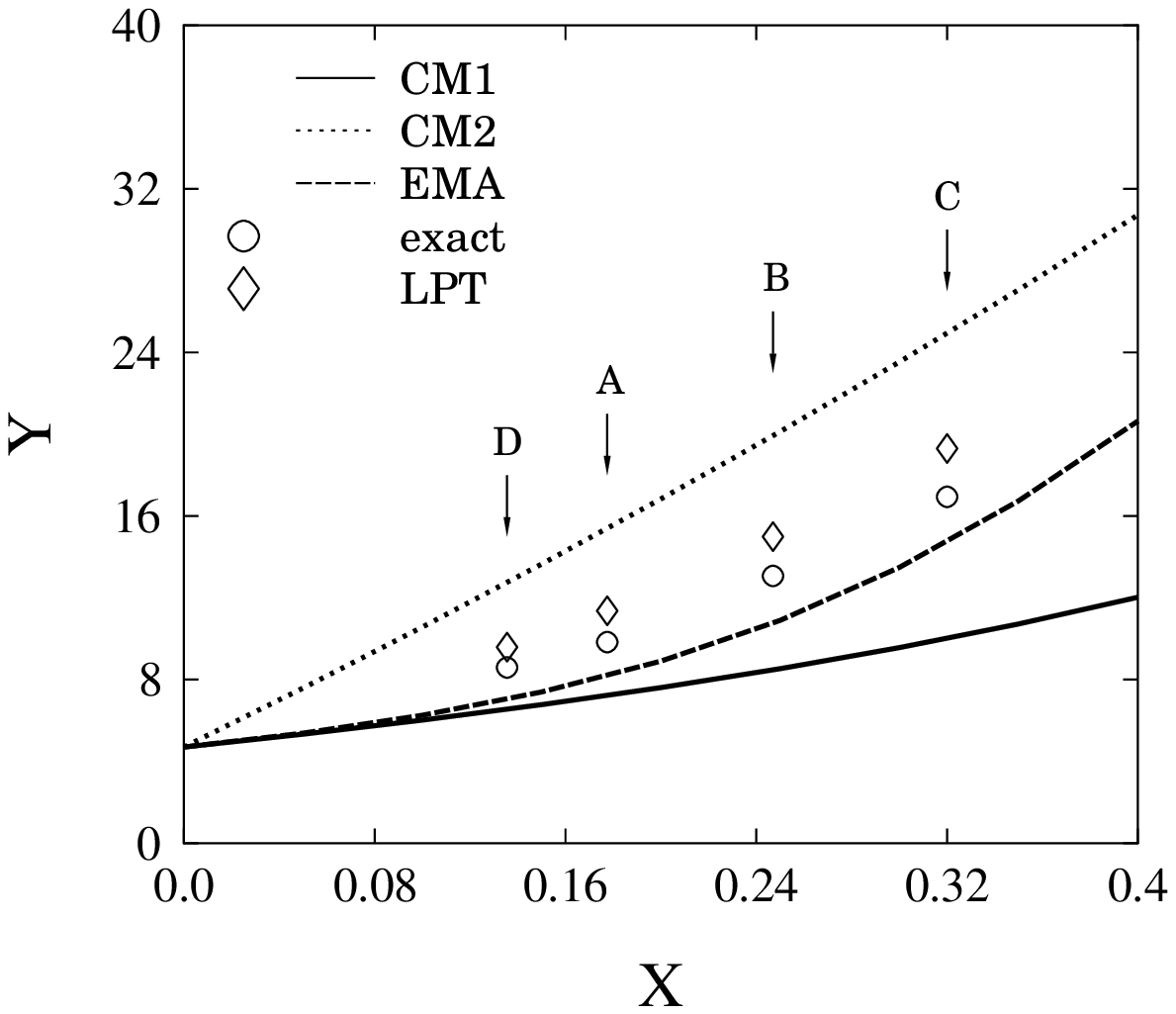,width=135mm}\\
Figure 2:
Comparison of approximate calculations for $\DKe$ evaluated
from eq. (\ref{epsc}) (upward triangles), eq.(\ref{epsb}) 
(downward triangles), eq.(\ref{EMA}) (squares), and
eq.(\ref{LPT1} (diamonds) with the exact result 
(circles).
\end{center}

Of course the comparison with $\DKe$ at zero frequency is not
sufficient to judge the quality of the approximations.
Solutions of the frequency dependent complex dielectric
function are in preparation \cite{WBH99}, and the simultaneous
comparison of real and imaginary parts is expected to provide 
further insight \cite{hil94g}.
Specifically, the solution of the frequency dependent 
inverse problem for local porosity theory is expected 
to yield information about porosity and connectivity fluctuations.
We emphasize, however, that first the use of $L_p$ must be
better established.
Currently we view it as a successful empirical rule based 
on the available data.
Further tests are necessary to corroborate or reject it.

ACKNOWLEDGEMENT: We are grateful to Dr. P.E. {\O}ren and Dr.S. Bakke 
for discussion, and for providing us with the experimental data sets.
We thank the Deutsche Forschungsgemeinschaft and the GKKS at the
Universit{\"a}t Stuttgart for financial support.


\begin{thebibliography}{10}

\bibitem{bot73}
C.~B{\"o}ttcher, {\em Theory of Electric Polarization}, vol.~I.
\newblock Amsterdam: Elsevier Scientific Publishing Co., 1973.

\bibitem{BB78}
C.~B{\"o}ttcher and P.~Bordewijk, {\em Theory of Electric Polarization},
  vol.~II.
\newblock Amsterdam: Elsevier Scientific Publishing Co., 1978.

\bibitem{kor84}
J.~Korringa, ``The influence of pore geometry on the dielectric properties of
  clean sandstones,'' {\em Geophysics}, vol.~49, p.~1760, 1984.

\bibitem{ken84}
W.~Kenyon, ``Texture effects on megahertz dielectric properties of calcite rock
  samples,'' {\em J. Appl. Phys.}, vol.~55, p.~3153, 1984.

\bibitem{KN87}
R.~Knight and A.~Nur, ``The dielectric constant of sandstones, 60 k{H}z to 4
  {MH}z,'' {\em Geophysics}, vol.~52, p.~644, 1987.

\bibitem{HN89}
I.~Holwech and B.~N{\o}st, ``Dielectric dispersion measurements of salt-water
  saturated porous glass,'' {\em Phys.Rev.B}, vol.~39, p.~12845, 1989.

\bibitem{TKS90}
M.~Taherian, W.~Kenyon, and K.~Safinya, ``Measurement of dielectric response of
  water saturated rocks,'' {\em Geophysics}, vol.~55, p.~1530, 1990.

\bibitem{has95}
E.~Haslund, ``Dielectric dispersion of salt water saturated porous glass
  containing thin glass plates,'' {\em Geophysics}, vol.~61, p.~722, 1996.

\bibitem{fri98}
S.~Friedman, ``A saturation-dependent composite sphere model for describing the
  effective dielectric constant of unsaturated porous media,'' {\em Water.
  Resources Res.}, vol.~34, p.~2949, 1998.

\bibitem{PNW78}
J.~Poley, J.~Nooteboom, and P.~de~Waal, ``Use of {VHF} dielectric measurements
  for borehole formation analysis,'' {\em The Log Analyst}, vol.~19, p.~8,
  1978.

\bibitem{HN85}
J.~Hearst and P.~Nelson, {\em Well Logging for Physical Properties}.
\newblock New York: McGraw-Hill, 1985.

\bibitem{hil91d}
R.~Hilfer, ``Geometric and dielectric characterization of porous media,'' {\em
  Phys. Rev. B}, vol.~44, p.~60, 1991.

\bibitem{WBH99}
J.~Widjajakusuma, B.~Biswal, and R.~Hilfer.
\newblock to be published.

\bibitem{HS62}
Z.~Hashin and S.~Shtrikman, ``A variational approach to the theory of effective
  magnetic permeability of multiphase materials,'' {\em J. Appl. Phys.},
  vol.~33, p.~3125, 1962.

\bibitem{ber82}
D.~Bergman, ``Rigorous bounds for the complex dielectric constant of a
  two-component composite,'' {\em Ann. Phys.}, vol.~138, p.~78, 1982.

\bibitem{tor91}
S.~Torquato, ``Random heterogeneous media: Microstructure and improved bounds
  on effective properties,'' {\em Applied mechanics reviews}, vol.~44, p.~37,
  1991.

\bibitem{hil92a}
R.~Hilfer, ``Local porosity theory for flow in porous media,'' {\em Phys. Rev.
  B}, vol.~45, p.~7115, 1992.

\bibitem{hil94b}
R.~Hilfer, B.N{\o}st, E.Haslund, Th.Kautzsch, B.Virgin, and B.D.Hansen, ``Local
  porosity theory for the frequency dependent dielectric function of porous
  rocks and polymer blends,'' {\em Physica A}, vol.~207, p.~19, 1994.

\bibitem{hil94g}
E.~Haslund, B.~Hansen, R.~Hilfer, and B.~N{\o}st, ``Measurement of local
  porosities and dielectric dispersion for a water saturated porous medium,''
  {\em J. Appl. Phys.}, vol.~76, p.~5473, 1994.

\bibitem{hil95d}
R.~Hilfer, ``Transport and relaxation phenomena in porous media,'' {\em
  Advances in Chemical Physics}, vol.~XCII, p.~299, 1996.

\bibitem{hil98a}
B.~Biswal, C.~Manwart, and R.~Hilfer, ``Threedimensional local porosity
  analysis of porous media,'' {\em Physica A}, vol.~255, p.~221, 1998.

\bibitem{hil99}
R.~Hilfer.
\newblock to be published.

\bibitem{hil92f}
F.~Boger, J.~Feder, R.~Hilfer, and T.~J{\o}ssang, ``Microstructural sensitivity
  of local porosity distributions,'' {\em Physica A}, vol.~187, p.~55, 1992.

\bibitem{hil96g}
C.~Andraud, A.~Beghdadi, E.~Haslund, R.~Hilfer, J.~Lafait, and B.~Virgin,
  ``Local entropy characterization of correlated random microstructures,'' {\em
  Physica A}, vol.~235, p.~307, 1997.

\bibitem{hil98h}
J.~Widjajakusuma, B.~Biswal, and R.~Hilfer, ``Quantitative prediction of
  effective material properties of heterogeneous media,'' {\em Comp. Mat.
  Sci.}, p.~in print, 1999.

\bibitem{BMHSO99}
B.~Biswal, C.~Manwart, R.~Hilfer, S.~Bakke, and P.E.{\O}ren, ``Quantitative
  analysis of experimental and synthetic microstructures for sedimentary
  rock.''
\newblock to be published.

\bibitem{hug96}
B.~Hughes, {\em Random Walks and Random Environments}, vol.~2.
\newblock Oxford: Clarendon Press, 1996.

\bibitem{bea72}
J.~Bear, {\em Dynamics of Fluids in Porous Media}.
\newblock New York: Elsevier Publ. Co., 1972.

\bibitem{SA92}
D.~Stauffer and A.~Aharony, {\em Introduction to Percolation Theory}.
\newblock London: Taylor and Francis, 1992.

\end{thebibliography}
\end{document}